# Measurement of the transfer function for a spoke cavity of C-ADS Injector I *

Xue-Fang Huang(黄雪芳)[1;2;1)] Yi Sun(孙毅)[2] Guang-Wei Wang (王光伟)[2] Shao-Zhe Wang (王少哲)[1;2]
Xiang Zheng(郑湘)[3] Qun-Yao Wang (王群要)[2] Rong Liu (刘熔) [2]Hai-Ying Lin (林海英)[2] Mu-Yuan Wang (王牧源) [1;2]

1 University of Chinese Academy of Sciences, Beijing 100039, China
2 Institute of High Energy Physics, CAS, Beijing 100049, China
3 Shanghai Institute of Applied Physics, CAS, Shanghai 201800, China

**Abstract**: The spoke cavities mounted in the China Accelerator Driven sub-critical System (C-ADS) have high quality factor(*Q*) and very small bandwidth, making them very sensitive to mechanical perturbations whether external or self-induced. The transfer function is used to characterize the response of the cavity eigen frequency to the perturbations. This paper describes a method to measure the transfer function of a spoke cavity. The measured Lorentz transfer function shows there are 206 Hz and 311 Hz mechanical eigenmodes excited by Lorentz force in the cavity of C-ADS, and the measured piezo fast tuner transfer function shows there are 12 mechanical eigenmodes from 0 to 500 Hz. According to these results, some effective measures have been taken to weaken the influence from helium pressure fluctuation, avoid mechanical resonances and improve the reliability of RF system.

## 1. Introduction

The China Accelerator Driven sub-critical System (C-ADS) is a pilot project to solve the nuclear waste problem [1]. The ADS injector I Proton Linac, independently designed and built by Institute of High Energy Physics, Chinese Academy of Sciences (IHEP), has successfully achieved guide specific objective on June 17, 2016. The proton beam peak current is above 10.03 mA and the output energy is 10.1 MeV. For the first time in the world, this injector uses 14 spoke superconducting radio frequency cavities with the extremely low β (β=0.12) [2-3], which are divided into two groups and installed into two cryomodules (CM1 and CM2), respectively. The main RF parameters of Spoke012 cavities are listed in Table1.

The high load quality factor $Q_L$ and narrow bandwidth as shown in Table 1 make the cavity extremely sensitive to any mechanical perturbations and thus make it difficult to control a spoke cavity. The perturbations detuning the cavity may origin from beam loading, helium pressure fluctuation, Lorentz force or microphonics [4-5]. Theoretically, a feedforward loop and a feedback loop with piezo fast tuner can be utilized to suppress the frequency detuning [6-7]. But how to practically realize these loops effectively is an ongoing topic. The transfer function is introduced to characterize the interaction between the RF field in cavity and the detuning forces, such as the Lorentz force and the force of a piezo acting on the cavity walls. With the measurement of transfer function, mechanical vibrations at certain frequencies can be studied in the cavity environment, thus the control scheme can be optimized so as to weaken the influence from helium pressure fluctuation, avoid mechanical resonances and improve the reliability of RF system. In this paper, a measurement method is introduced to determine the Lorentz transfer function and the piezo fast tuner transfer function of a spoke cavity mounted in C-ADS, and the measurement results will be also discussed.

Table 1. Main RF parameters of Spoke012 cavities.

| Parameter | Design | Measured |
|---|---|---|
| Frequency(MHz) | 325 | 324.833 |
| $Q_L$ | $7.5\times10^5$ | $>4\times10^5$ |
| operating mode | CW | CW |
| Operating temperature(K) | 4.2 or 2 | 2 |
| Beam current (pulse) (mA) | 10 | 10.03 |
| *Eacc* (in operation)( MV/m) | 6.08 | 5~8 |
| Bandwidth (±3 dB)(Hz) | 433 | 400~ 800 |
| Cavity Tuning Sensitivity(kHz/um) | 1 | 1.1 |
| Factor of static LFD (Hz/(MV/m)^2) | -4 | -15~-5 |

## 2. Principle of the measurement

The Lorentz force, generated by the interaction between the electromagnetic field inside the cavity and the cavity wall current, deforms the cavity shape and subsequently shifts the cavity resonance frequency. The Lorentz transfer function characterizes the coupling between the mechanical eigenmodes driven by the Lorentz force and the RF field in cavity [8]. If the mechanical eigenmodes are indexed by ν, then the RF frequency shift driven by mechanical forces $F_\nu$ satisfies the relation as follows [9]:

$$\frac{\mathrm{d}^2\Delta\omega_\nu}{dt^2}+\frac{2\Omega_\nu}{Q_\nu}\frac{\mathrm{d}\Delta\omega_\nu}{dt}+\Omega_\nu^2\Delta\omega_\nu=-\frac{\omega\Omega_\nu^2F_\nu^2}{c_\nu W}, \tag{1}$$

where $c_\upsilon$ is the elastic constant, $\Omega_\upsilon$ is the mechanical eigenfrequency, $Q_\upsilon$ is the quality factor, $W$ is the stored energy and $\Delta\omega_\upsilon$ is the frequency shift because of the mechanical eigenmode, and total frequency shift $\Delta\omega$ of the excited mechanical eigenmodes is:

$$\Delta\omega = \sum_\nu \Delta\omega_\nu. \tag{2}$$

*Supported by Proton linac accelerator I of China Accelerator Driven sub-critical System (Y12C32W129)
1) E-mail: huangxf@ihep.ac.cn

According to the Maxwell's equations, the mechanical forces $F_\nu$ on cavity walls and the stored energy $W$ of the electromagnetic field are proportional to the square of the accelerating gradient.

$$F_\nu = f_\nu \cdot E_{acc}^2, \tag{3}$$

$$W = W_0 \cdot E_{acc}^2. \tag{4}$$

Then equation (1) can be written as:

$$\frac{\mathrm{d}^2\Delta\omega_\nu}{dt^2}+\frac{2\Omega_\nu}{Q_\nu}\frac{\mathrm{d}\Delta\omega_\nu}{dt}+\Omega_\nu^2\Delta\omega_\nu=-\frac{\omega\Omega_\nu^2 f_\nu^2}{c_\nu W_0}E_{acc}^2, \tag{5}$$

and in the frequency domain is:

$$\Delta\omega=\sum_\nu\frac{-\omega\Omega_\nu^2 f_\nu^2 / c_\nu W_0}{\left(\Omega_\nu^2-\Omega^2\right)+2i\Omega\Omega_\nu / Q_\nu} * E_{acc}^2(\Omega)=2\pi\cdot\left(\sum_\nu k_\nu(\Omega)\right)\cdot E_{acc}^2(\Omega). \tag{6}$$

where the Lorentz transfer function (LTF) is:

$$LTF = \sum_\nu k_\nu(\Omega). \tag{7}$$

Similarly, the piezo fast tuner transfer function characterizes the coupling between the mechanical eigenmodes driven by the force of the piezo acting on the cavity walls and the RF field in cavity. Here, the force comes from the mechanical shift, which depends on the driving voltage forced on the piezo. So the driving voltage $U$ is used to indicate the magnitude of the force on the cavity walls. The relationship between the frequency shift and the driving voltage $U$ in frequency domain can be shown as:

$$\Delta\omega = 2\pi\cdot\left(\sum_\nu p_\nu(\Omega)\right)\cdot \mathrm{U}(\Omega), \tag{8}$$

where the piezo fast tuner transfer function (PTF) is:

$$PTF = \sum_\nu p_\nu(\Omega). \tag{9}$$

In electro-mechanical interactions, the total RF frequency variation can be regarded as the sum of the equations (6) and (8) (in this situation, equation (8) indicates the frequency variation caused by both the fast tuner and the slow tuner).

It can be shown from equation (6) that the Lorentz transfer function establishes the contact of the amplitude modulation of the cavity RF field and the frequency modulation driven by the corresponding Lorentz force. While equation (8) indicates that the piezo fast tuner transfer function establishes the contact of the frequency of the piezo driving voltage and the amplitude and phase of frequency modulation of the RF field in cavity. Both of the two kinds of transfer functions need to record the frequency change of the RF field in cavity. The difference is that what makes the change: for the former, it is the amplitude modulation of the RF field; for the latter, it is the amplitude modulation of the driving voltage on the piezo.

## 3. The measurement of the transfer function

The 7# spoke cavity in CM1 is chosen as the measurement object, and the measuring block diagram is shown in Fig.1. A 324.833 MHz reference (REF) sinusoid is generated by a signal source, and then is fed into a clock distribution IC AD9510 to generate IF signal and working clocks for ADC, DAC and FPGA. For this application, the IF is running at one-sixteenth times reference, while the AD/DA sampling clocks are both a quarter of reference. The IF signal is mixed with the REF to generate a LO signal which is used to down-convert the concerned RF signals to IF band and also to up-convert processed IF signals to RF band.

In the center of measurement system，one Stratix II EP2S60 FPGA is hosted on the FPGA board to implement the signal processing and data transmission. Three high-speed ADCs (LTC2255, ADC1/2/3 in Fig.1) are used to sample the reference (REF), the pick-up signal (Pt) and the incident signal (Pf). These ADCs have 14 bit accuracy, 125 MPS maximum sampling rate and 72.4 dB SNR. The sampling data from ADCs is converted to baseband I/Q utilizing digital IQ demodulator, the I/Q sequences are passed to a PI controller implemented in FPGA, and the PI outputs are used to construct digital waveform taking advantage of numerically controlled oscillator (NCO). Another NCO is also implemented to generate a modulation signal that can be used to module the generated digital waveform or sent to piezo controller directly by DAC3, which is a voltage-output DAC (LTC2630). The parameters of this NCO, I_set and Q_set, are set by user from communication ports. One high-speed dual DAC (ISL5927, DAC1/2 in Fig.1) is used to convert the digital waveform to analog waveform. The digital waveform without modulation is converted by the DAC1, while the modulated is converted by the DAC2. It depends on a RF switch whose output between DAC1 and DAC2 will be sent to the solid-state amplifier (SSA). The output of DAC3 ranges from 0 V to 4.096 V, and the gain of the piezo controller is 100, so the bias voltage is set at 204.8 V to keep the piezo in a safe working condition. In the actual measurement, the DAC3 can also act as a low pass filter by attenuating the amplitude of the output signal when the output frequency increases.

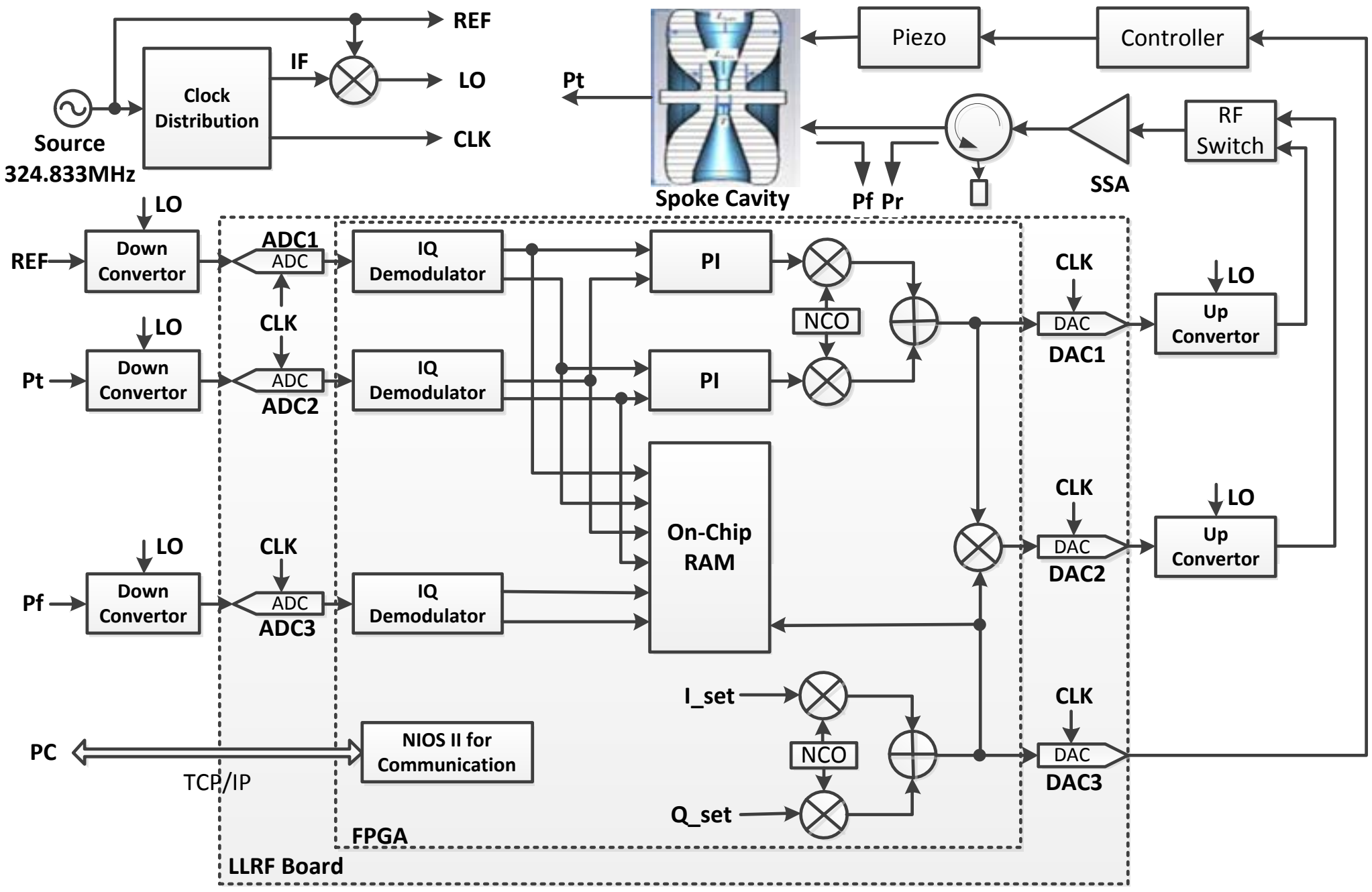

Fig. 1. Block diagram of the measuring system.

The IQ pairs of REF, Pt, Pf and the piezo driving signal are buffered with FPGA on-chip RAMs. The buffering rate is 4096 Hz, and the buffering duration is 2 seconds each frequency point. The buffered data is transmitted to the host PC by 10M Ethernet port mounted on the FPGA board, and is finally recorded in EXCEL format by a Labview program on host PC.

The method to determine the Lorentz transfer function is to measure the relationship between the RF driving signal Pf and the cavity output Pt at different modulation frequency. During operation, the RF frequency is fixed, and the SSA is driven by DAC2 so the amplitude and phase of Pf can be modulated by setting different I_set, Q_set. The amplitude and phase of the modulated RF driving signal are recorded by ADC3, while that of the RF field in cavity are measured by ADC2. The modulation frequency ranges from 0 to 1000 Hz in 1Hz increment, and the relationship between Pf and Pt is recorded at each frequency.

Similarly, the method to determine the piezo fast tuner transfer function is to measure the amplitude and phase error seen in the cavity output signal when the piezo driving voltage is modulated. The piezo has the effect of shifting the cavity resonant frequency that results in a significant error in amplitude and phase in Pt when operating at the frequencies near cavity resonant frequency. During operation, the SSA is driven by DAC1, and the piezo controller is driven by DAC3. The vibration frequency of the piezo driving voltage also ranges from 0 to 1000 Hz in 1Hz increment, and the relationship between Pt and piezo driving voltage is recorded at each frequency.

To reduce the influence of microphonics, the measurement was carried out in the midnight when external disturbances are relatively small. Besides, the amplitude of the modulation signal is large enough to make the frequency shift mainly caused by modulation while the shift by microphonics is insignificant.

## 4. Results and Discussions

### 4.1 The result of Lorentz transfer function

The Lorentz transfer function of the spoke cavity in CM1 has been measured when the accelerating gradient is 3MV/m, and the result is plotted in Fig.2. The accelerating gradient should be large enough because the modulation amplitude can be larger than 2MV/m. The horizontal axis is the vibration frequency of the amplitude modulation signal. The vertical axis of the top half graph is the amplitude of the frequency modulation, which is forced on the cavity RF, divided by $\Delta E^2$. $\Delta E^2(\Omega) = E^2acc,max(\Omega) - E^2acc,min(\Omega)$ , is the difference between the squared maximum and the squared minimum of the accelerating gradient sinusoidal amplitude modulation. The vertical axis of the lower half graph gives the relative phase difference between the amplitude modulation signal of the RF field and the frequency modulation of the cavity.

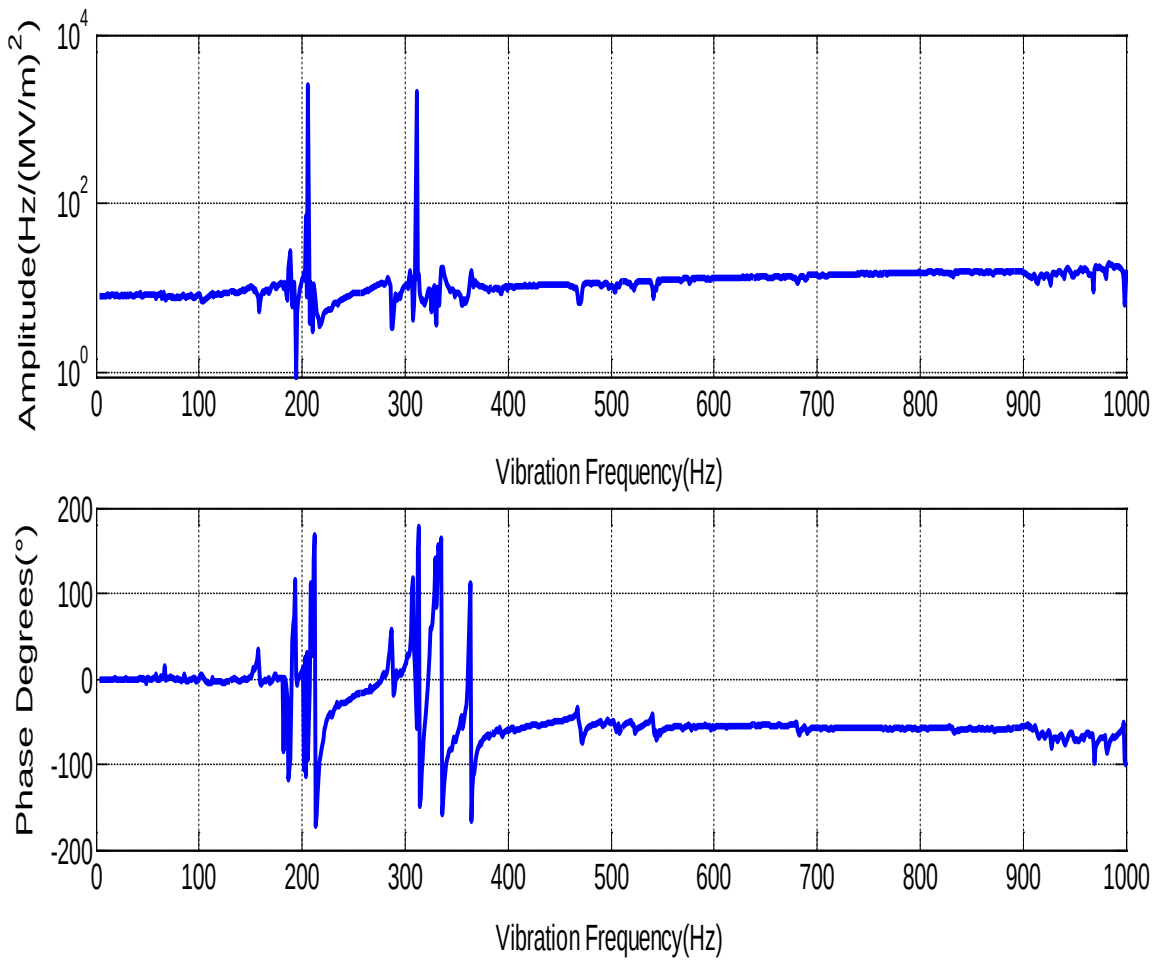

Fig. 2. Measurement result of the Lorentz transfer function for 7# cavity in CM1.

It is shown from the result that there are two mechanical eigenmodes driven by the Lorentz force at the frequency point of 206 Hz and 311 Hz in cavity. The two mechanical eigenmodes' *Q* values are so large that the amplitude and the phase near them have drastic changes, leading to detectable side bands in the cavity probe signal Pt. The measured spectrogram of the pick-up signal of the 7# Spoke012 cavity shown in Fig.3 indicates that under the open loop state, the sideband of 311 Hz can be driven rather easily, so in actual operation, it should be careful to avoid driving these two mechanical resonances. It is also shown in Fig.2 that there is no mechanical eigenmode when the vibration frequency is lower than 185 Hz, i.e., the Lorentz force doesn't couple with the microphonics in low frequency.

According to the Lorentz transfer function, several proposals can be offered to eliminate or allay the influence of these modes, since the cavities and cryostats are already in place. On the one hand, vibration sources whose resonant frequency is near 206 Hz or 311 Hz should be far away from the accelerator tunnel. At the same time, quarantine measures are necessary to keep away from coupling with these frequencies. On the other hand, an effective solution, based on mechanical damping, is proposed to suppress these dangerous modes.

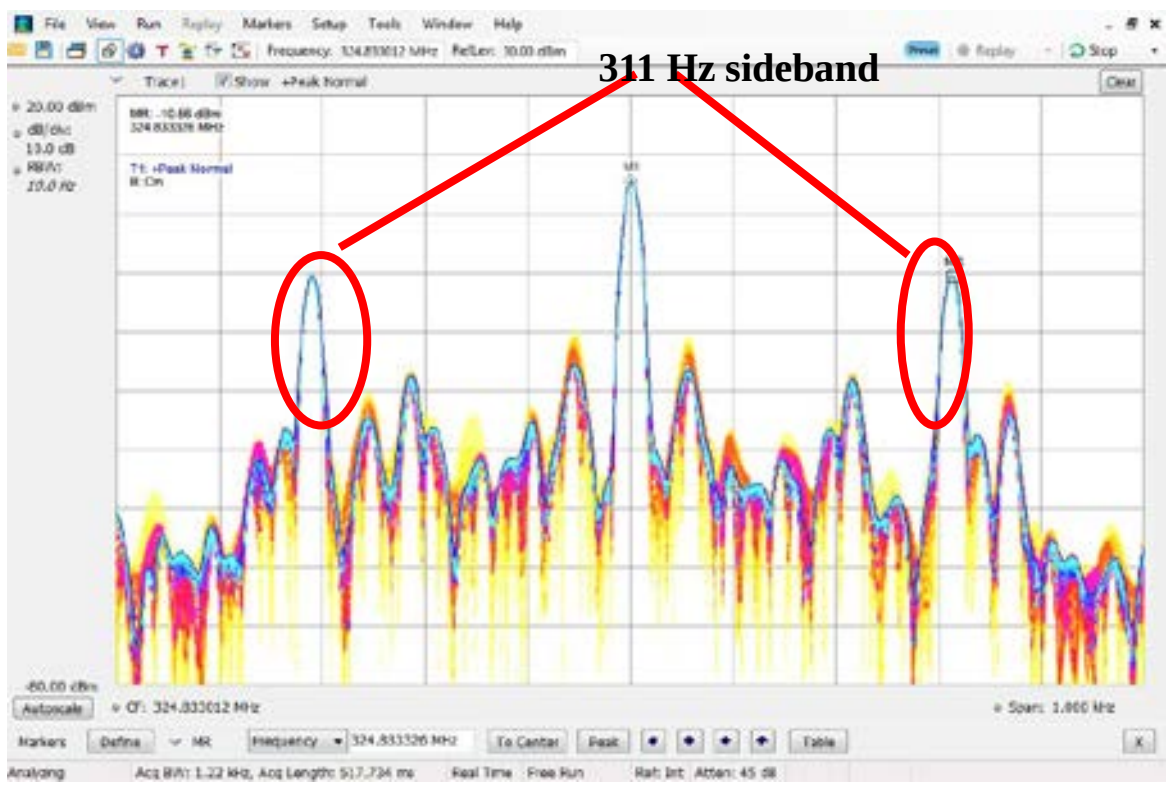

Fig. 3. Spectrogram of the pick-up signal of the Spoke012 cavity on working, which shows that the 311 Hz sideband has been driven.

### 4.2 The result of piezo fast tuner transfer function

The piezo fast tuner transfer function of the 7# Spoke012 cavity in CM1 has been measured when the accelerating gradient is 5 MV/m, and is plotted in Fig.4. The horizontal axis is the amplitude modulation frequency of the piezo driving voltage. The top half of the graph is the amplitude of the frequency modulation for the RF field in cavity divided by the amplitude of piezo driving voltage amplitude modulation signal. The lower half of the plot is the relative phase difference between the frequency modulation signal of the cavity and the amplitude modulation signal of the piezo driving voltage.

The result shows that in low frequency (0-30 Hz), there is no mechanical eigenmode driven by piezo. But up to 50 Hz vibration frequency, the motion of the piezo has obvious coupling with a mechanical eigenmode. In the range from 300 to 400 Hz, there are larger couplings with mechanical eigenmodes. Especially in 315 Hz, the largest amplitude reaches 9.9 Hz/V. The calculated results of the top-twelve eigenmodes driven by piezoelectric actuators are plotted in Fig. 5.

The measurement of the transfer function between the piezo drive signal and the cavity deformation is the key to design the piezo-based control loop. The result indicates that band rejection filters from 300 Hz to 400 Hz on the piezo driving voltage should be added to avoid the mechanical resonance. It is also the factual basis to design an appropriate low pass filter of 30 Hz in the frequency-tuning loop of this cavity. The frequency-tuning loop with the low pass filter works very well. It suppresses the low frequency variation resulted from the helium pressure fluctuation and other causes on the maximum limit instead of exciting the mechanical resonances.

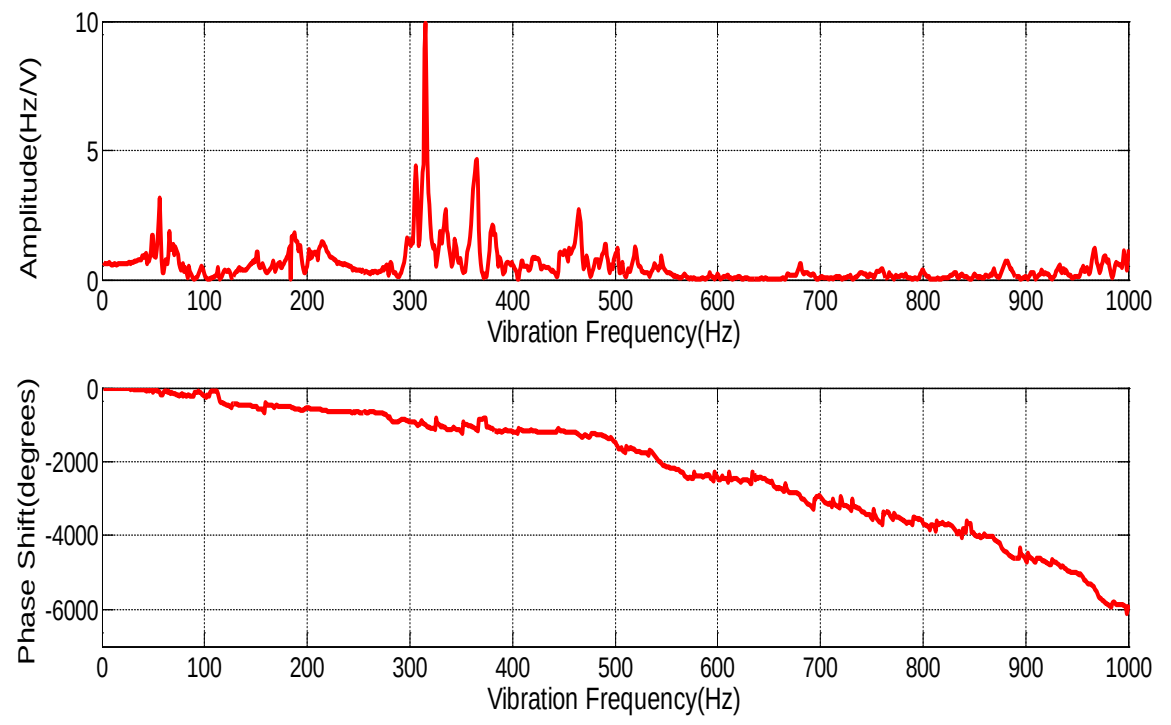

Fig. 4. Measurement result of the fast tuner transfer function for 7# cavity in CM1.

## 5. Summary

A first measurement of the Lorentz transfer function and the piezo fast tuner transfer function of a spoke012 cavity in C-ADS has been shown in this paper. The measured Lorentz transfer function shows 206 Hz and 311 Hz sidebands can be driven easily when the cavity accelerating gradient is raised, which is consistent with the actual operation result measured by an spectrum analyzer. It suggests that certain vibration sources should be far away from the cavities or several quarantine measures are needed. Besides, an effective solution, based on mechanical damping, is proposed to supress these dangerous modes. The measured piezo fast tuner transfer function shows there are many mechanical eignmodes from 0 Hz to 500 Hz, either in the transverse or in the vertical direction. It is a very significant to design piezo-based control loop. According to the result, a 30 Hz low-pass filter was added in the piezo control loop to suppress the frequency variation without exciting the mechanical resonances. Both transfer functions can be used to predict the frequency shift caused by the Lorentz force or piezo fast tuner, and these results will play an important role in the coming feedforward control system to suppress disturbances.

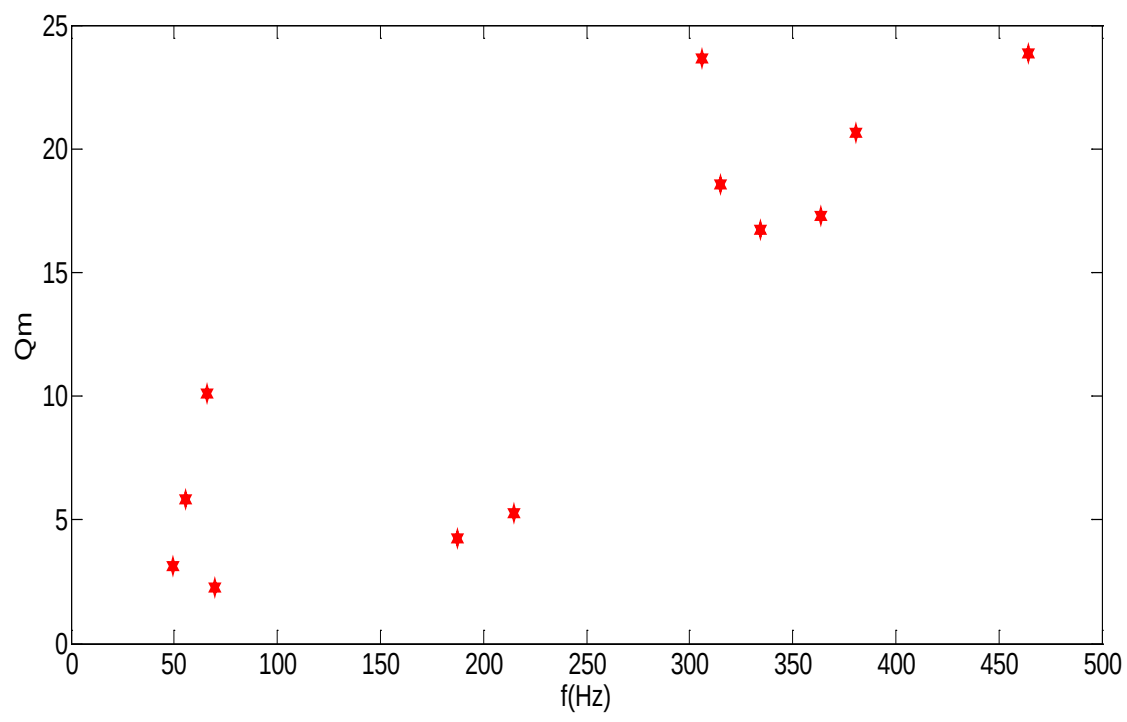

Fig. 5. Top-twelve eigenmodes driven by piezoelectric actuators (7# Spoke012 cavity in CM1).

## References

[1] J. Y. Tang, P. Cheng, H. P. Geng et al, Conceptual Physics Design for the China-ADS Linac, in Proceedings of PAC2013, Pasadena, CA, USA

[2] H. Li, P. Sha, H. Huang et al, Chin. Phys. C, **36**(8): 761-764 (2012)

[3] H. Li, J. P. Dai, P. Sha et al, Chin. Phys. C, **38**(7): 130-134 (2014)

[4] N. Liu, Y. Sun, G. W. Wang et al, Tuner control system of spoke012 SRF cavity for C-ADS injector I at IHEP, Chin. Phys. C, to be published

[5]http://inspirehep.net/record/1416561/files/fermilab-conf-15-616-td.pdf , retrieved 22th January 2016

[6] Z.A. Conway and M.U. Liepe, FAST PIEZOELECTRIC ACTUATOR CONTROL OF MICROPHONICS IN THE CW CORNELL ERL INJECTOR CRYOMODULE, in Proceedings of PAC09, Vancouver, BC, Canada

[7] R. Rybaniec, L. J. Opalski, ISE WUT et al, MICROPHONIC DISTURBANCES PREDICTION AND COMPENSATIONIN PULSED SUPERCONDUCTING ACCELERATORS, in Proceedings of IPAC2015, Richmond, VA, USA

[8]http://www.diva-portal.org/smash/get/diva2:716380/FULLTEXT01.pdf, retrieved May 2014

[9] Zachary A. Conway, Electro-mechanical interacttions in superconducting spoke-loaded cavities, Ph.D. Thesis (Urbana: University of Illinois at Urbana-Champaign, 2007)